\documentclass[prl,aps,twocolumn,showpacs,preprintnumbers,amsmath,amssymb]{revtex4}

\usepackage{graphicx}
\usepackage{dcolumn}

\begin{document}

\title{Phonon softening and Superconductivity triggered by spin-orbit coupling
in simple-cubic $\alpha$-polonium}

\author{Chang-Jong Kang, Kyoo Kim, B. I. Min}
\affiliation{Department of Physics, PCTP,
	Pohang University of Science and Technology, Pohang 790-784, Korea}
\date{\today}

\begin{abstract}
We have investigated the mechanism of stabilizing the simple-cubic ($sc$)
structure in polonium ($\alpha$-Po),
based on the phonon dispersion calculations using the
first-principles all-electron band method.
We have demonstrated that the stable sc structure results
from the suppression of
the Peierls instability due to the strong spin-orbit
coupling (SOC) in $\alpha$-Po.
We have also discussed the structural chirality realized in $\beta$-Po,
as a consequence of the phonon instability.
Further, we have explored the possible superconductivity in $\alpha$-Po,
and predicted that it becomes a superconductor with $T_{c} \sim 4$ K.
The transverse soft phonon mode at $\bf q \approx \frac{2}{3}$R,
which is greatly influenced by the SOC, plays an important role
both in the structural stability and the superconductivity in $\alpha$-Po.
\end{abstract}


\maketitle

Polonium (Po), which belongs to chalcogen group in the periodic table,
is unique in that it crystallizes in the simple-cubic (sc)
structure \cite{Beamer46}.
Po having atomic number Z = 84 was discovered by Marie and Pierre Curie
in 1898.
Po exists in two metallic allotropes, $\alpha$ and $\beta$-Po.
It is $\alpha$-Po that has a sc structure.
Above 348 K, $\alpha$-Po transforms
to $\beta$-Po, which has the trigonal structure \cite{Maxwell49}.
Selenium (Se) and Tellurium (Te), which are isoelectronic elements to Po
in the chalcogen group, also have the trigonal spiral structure \cite{Beister90}.
Note that the trigonal structure can be derived from the $sc$ structure
by elongation or contraction along the [111] direction \cite{Kresse94},
which is known to occur due to the Peierls distortion
in $p$-bonded systems of Se and Te \cite{Burdett83}.

Recently, there have been several reports to explore
the origin of the stabilized $sc$ structure in
Po \cite{Min06,Legut07,Min09,Legut09,Legut10,Verst10}.
General consensus so far is that the large relativistic effects in Po
play an important role.
The Peierls instability tends to be
suppressed in Po by the relativistic effects.
However, there was controversy on the role of spin-orbit coupling (SOC)
of $6p$ electrons. Min {\it et al.}\cite{Min06} claimed that
the SOC in addition to the scalar-relativistic (SR) effects of mass-velocity
and Darwin terms is essential, while Legut {\it et al.}\cite{Legut07,Legut09}
claimed that the SR effects are already enough to stabilize $sc$-Po.
The refined and comprehensive band calculations by the former group,
by considering all the relativistic corrections and different
exchange-correlation functionals
show that the SOC of $6p$ states is really important in
stabilizing the $sc$ phase of Po \cite{Min09}.

In general, the structural transition is induced by the
phonon softening instability,
and so the phonon dispersion calculations would provide
evidence of the structural stability.
Verstraete\cite{Verst10} calculated the phonon dispersion
of $sc$-Po based on the pseudo-potential band method, and showed that
the phonon softening instability does not occur even without including the SOC.
This result seems to support Legut {\it et al.}'s claim\cite{Legut07}
that the SR terms are sufficient to stabilize the $sc$ phase.
Thereafter, a couple of more studies were reported on
the phonon dispersion of the $\alpha$-Po at the ambient
pressure\cite{Belabbes10} and under the pressure\cite{Zaoui11},
by using the pseudo-potential band method.
Since Po has a tiny energy difference, order of 1 meV, between the $sc$ and
trigonal structures\cite{Min09},
special caution is needed to calculate and analyze the phonon dispersions.

The structural energetics in Po is known to be very sensitive
to the volume and the utilized exchange-correlation method in
the band calculation\cite{Min09,Legut10}.
The lattice constant $a_{exp}$ of $\alpha$-Po was first measured by
Beamer {\it et al.}\cite{Beamer49} to be 3.345 ${\AA}$.
Later, Desando {\it et al.}\cite{Desando66} reported
the refined lattice constant of $\alpha$-Po to be 3.359 ${\AA}$.
In the above band and phonon dispersion calculations,
the old $a^{old}_{exp}=3.345$ ${\AA}$ was referenced.
In the present study,  we have taken $a_{exp}=3.359$ ${\AA}$ as a reference.
We have found that the small difference in $a_{exp}$ is important
in analyzing the structural stability.

As mentioned above, the Peierls instability is
the key physics in Po \cite{Min06,Verst10,Legut10,Belabbes10}.
Even if the Peierls instability is suppressed by the SOC,
$\alpha$-Po in the $sc$ phase would still have the
soft phonon mode and so the
strong electron-phonon (EP) coupling.
Then the natural question is whether $\alpha$-Po would have
superconductivity or not.
To our knowledge, superconductivity in Po has not been explored yet,
probably due to its radioactive and toxic nature.

In this Letter, we have investigated phonon and superconducting properties
of $\alpha$-Po, employing the first principles all-electron band methods.
The phonon dispersions and superconducting properties of $\alpha$-Po
are studied with and without the SOC at different volumes.
By means of the phonon dispersion calculations,
we have explicitly demonstrated
that the stability of the sc-phase arises from the SOC,
which concludes the longstanding dispute
on the origin of stable sc structure of Po.
Further, we have predicted that $\alpha$-Po
is a superconductor with $T_{c}$ of $\sim 4$ K.

We have employed the FLAPW band method\cite{Freeman}, implemented
in the Elk package \cite{Elk}.
The SOC was included in second-variational scheme
and the local-density approximation (LDA) was used
for the exchange-correlation energy.
Actually, the choice of the LDA is more stringent test
for the phonon stability than that of the
generalized gradient approximation (GGA),
because the GGA always gives more softened phonons than the LDA.
For $sc$-Po, $a^0_{th}=3.335 {\AA}$ was obtained
in the LDA+SOC scheme\cite{Band}, which is
in good agreement with $a_{exp}=3.359 {\AA}$ of $\alpha$-Po.
Phonon dispersions were obtained by using the supercell method,
implemented in the Elk package.
The force constants and the dynamical matrix are obtained
from the Hellmann-Feynman forces calculated
with small individual displacements of nonequivalent atoms \cite{Yu91}.

\begin{figure}[t]
\includegraphics[width=7.5 cm]{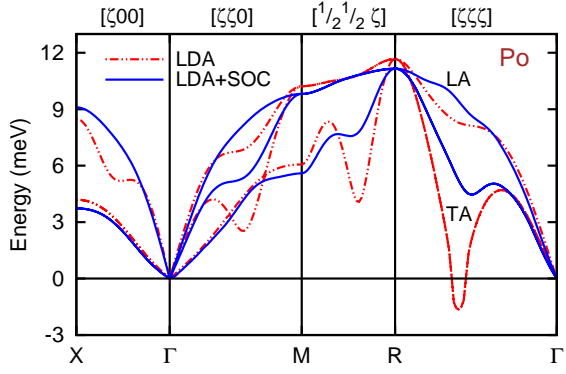}
\caption{(Color Online) Phonon dispersions of $sc$-Po at $a=a_{exp}$
with SOC (LDA+SOC: blue-solid line) and without it (LDA: red-dotted line).
The imaginary phonon frequencies are represented as negative values
in the figure.}
\label{ph-po1}
\end{figure}

\begin{figure}[t]
\includegraphics[width=8.5 cm]{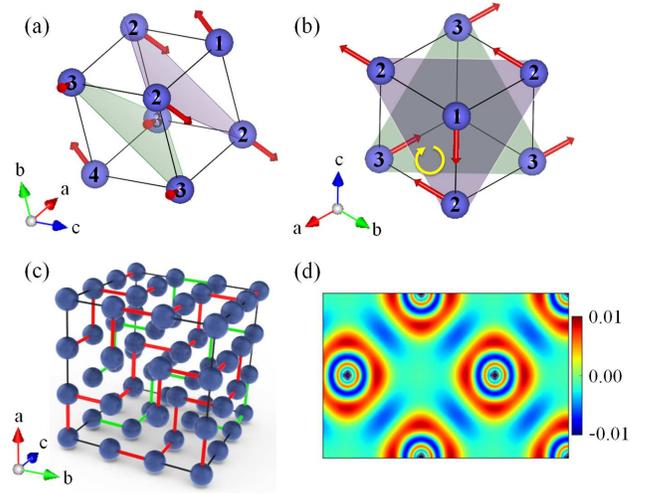}
\caption{(Color Online)
(a) One of two degenerate normal mode at $\bf q \approx \frac{2}{3}$R in $sc$-Po.
All the displacement vectors are placed on (111) planes.
(b) The same normal mode shown along the [111] viewpoint,
which manifests a clockwise helicity.
(c) The helical chain structure of $\beta$-Po with definite chirality.
Red and Green lines represent short bonds.
(d) The difference of charge density of $sc$-Po
in the LDA+SOC and that in the LDA (in unit of $e/{\AA}^3$).
Notice the charge density depletion in-between neighboring ions.
}
\label{mode}
\end{figure}

Figure~\ref{ph-po1} shows the phonon dispersions of
$\alpha$-Po at $a=a_{exp}$
with and without the SOC \cite{Phonon}.
In the LDA scheme without the SOC
(corresponding to the SR band scheme),
the phonon softenings occur along all the high symmetry
directions. Among those, phonon frequency along $\Gamma$-R
($\bf q \approx \frac{2}{3}$R)
is imaginary, which indicates the structural distortion
along the [111] direction \cite{GGA}.
In contrast, in the LDA+SOC scheme
(corresponding to the fully-relativistic band scheme),
the phonons become
hardened and the imaginary phonon softening disappears.
These features reflect that, without the $6p$ SOC,
the $sc$ structure of $\alpha$-Po is unstable to a trigonal structure,
as in Se and Te.

\begin{figure}[t]
\includegraphics[width=7.5 cm]{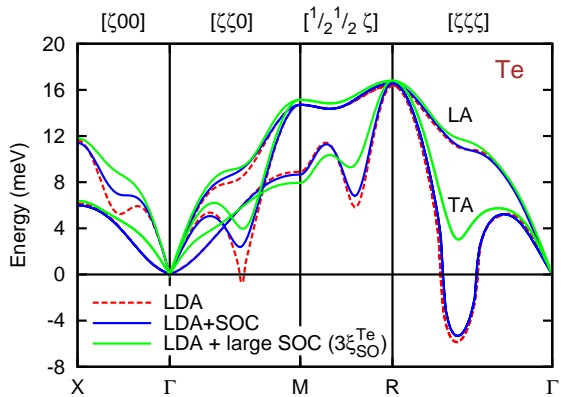}
\caption{(Color Online) Phonon dispersion curves of $sc$-Te
at the lattice constant of 3.210 ${\AA}$,
without the SOC (LDA), with the SOC (LDA+SOC),
and with the three times larger SOC (3$\xi^{Te}_{SO}$).
}
\label{ph-sc-Te}
\end{figure}

Note that, for the softened phonon at $\bf q \approx \frac{2}{3}$R,
there are two degenerate normal modes having opposite helicities.
The one has a clockwise helicity as shown in Fig.~\ref{mode}(b),
while the other has a counterclockwise helicity.
Thus the softened phonon at $\bf q \approx \frac{2}{3}$R
induces the structural transformation to the trigonal $\beta$-Po,
which has the helical chain structure with definite chirality
(Fig.~\ref{mode}(c)).
The chiral structure realized in $\beta$-Po was once proposed
in Te~\cite{Fukutome84}, which was recently explained in terms of the
orbital-ordered chiral charge-density wave (CDW)~\cite{Wezel10,Wezel11}.
This orbital-ordered CDW would be suppressed, if the SOC were larger,
as will be discussed in Fig.~\ref{ph-sc-Te}.
The chirality of similar feature was also observed
in $1T$-TiSe$_{2}$ \cite{Ishioka10}.
Interestingly, both Te and $1T$-TiSe$_{2}$ exhibit superconductivity
under high pressure \cite{Akahama92,Mauri96}
and Cu intercalation \cite{Morosan06}, respectively,
suggesting that the superconductivity emerges when the CDW is suppressed.

The phonon softening in Fig.~\ref{ph-po1} is closely related to the
Peierls mechanism.
The phonon softening is described by the renormalization of
phonon frequency by the EP interaction (the so-called Kohn anomaly),
\begin{equation}
\omega^2({\bf q})=\Omega^2({\bf q})-
|\tilde{g}_{ep}({\bf q})|^2\chi_{0}({\bf q}),
\end{equation}
where $\omega({\bf q})$ and $\Omega({\bf q})$ correspond to
renormalized and bare phonon frequencies, respectively,
and $\tilde{g}_{ep}$ and $\chi_{0}({\bf q})$ are
the EP coupling parameter and the charge susceptibility.
Min {\it et al.}\cite{Min06} obtained that $\chi_{0}({\bf q})$
in $\alpha$-Po has the highest peak at $\bf q \approx \frac{2}{3}$R
due to the Fermi surface nesting.
They also showed that the SOC suppresses the Fermi surface nesting effect
and accordingly the $\chi_{0}({\bf q})$ value,
whereby they predicted that the phonon softening
is weakened and finally the imaginary phonon softening disappears.

The suppression of the Peierls instability by the SOC
occurs not only through the mechanism of Eq. (1) but also
through weakening of the bonding strength.
This feature is shown clearly in Fig.~\ref{mode}(d), which plots the difference
of the charge density in the LDA+SOC and that in the LDA.
It is evident that, due to the SOC, the charge density is depleted
in-between the neighboring ions, which results in the weakening
of the directional bondings of Po chains.

\begin{figure}[t]
\includegraphics[width=7.5 cm]{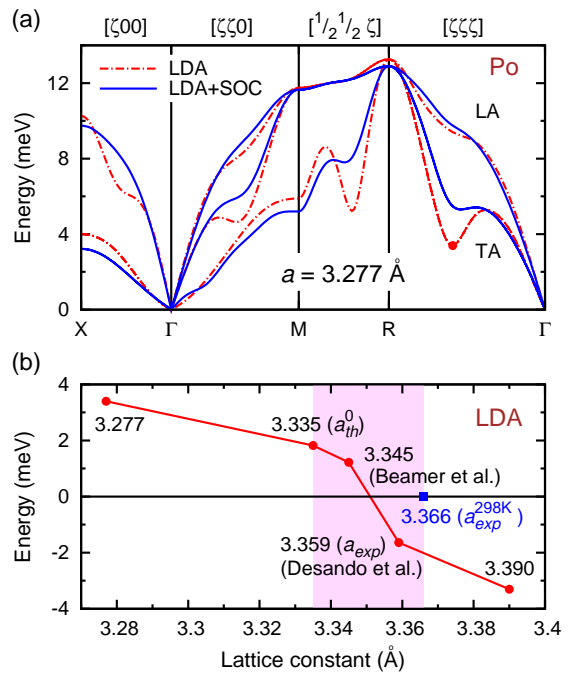}
\caption{(Color Online)
(a) The phonon dispersion curves of the $sc$-Po
with and without the SOC at the reduced volume $V/V_{exp} = 0.93$
($a = 3.277\AA$).
The filled circle corresponds to the phonon mode at
$\bf q \approx \frac{2}{3}$R.
The overall phonon modes are hardened at the reduced volume.
(b) The behavior of softened TA phonon mode
at $\bf q \approx \frac{2}{3}$R in the LDA
with varying the lattice constant.
Here, $a = 3.335\AA$ ($V/V_{exp} = 0.98$) corresponds to $a^0_{th}$
of the LDA+SOC, while $a = 3.345\AA$ and $a = 3.359\AA$
correspond to the low temperature experimental lattice constants
reported by Beamer et al.~\cite{Beamer49} and
by Desando et al.~\cite{Desando66}, respectively.
$a = 3.366\AA$ (filled square) corresponds to the lattice constant
at 298 K~\cite{Lide}.
}
\label{ph-po2}
\end{figure}

To verify the role of SOC in the structural stability more convincingly,
we have examined the phonon dispersions of Te
in the hypothetical $sc$ structure ($sc$-Te)
with artificially varying the strength of SOC of Te ($\xi^{Te}_{SO}$).
Figure~\ref{ph-sc-Te} shows that, in the LDA,
two imaginary phonon softenings occur along
the $\Gamma$-M and $\Gamma$-R directions,
while, in the LDA+SOC,
that remains only along the $\Gamma$-R direction.
However, when we increases the SOC three times ($3\xi^{Te}_{SO}$),
even the last imaginary phonon softening disappears eventually.
This feature reveals that Te would also have a $sc$ structure,
if the SOC of Te becomes larger.
The situation of the three times larger SOC
in Fig.~\ref{ph-sc-Te} actually simulates the case in Po,
since the SOC of $6p$ electrons in Po ($\xi^{Po}_{SO}= 1.90$ eV)
is almost three times stronger than that
of $5p$ electrons in Te ($\xi^{Te}_{SO}= 0.72$ eV).
The behavior in Fig.~\ref{ph-sc-Te} provide definite evidence
that the SOC really plays a role of suppressing the Peierls instability.

At the reduced volume, phonons tend to be hardened,
and so it is expected that the phonon softening observed in
Fig.~\ref{ph-po1} would be reduced.
Indeed, the phonon dispersions of $sc$-Po
at $a = 3.277\AA$ ($V/V_{exp} = 0.93$)
in Fig.~\ref{ph-po2}(a) do not exhibit
the imaginary phonon softenings both in the LDA and LDA+SOC schemes.
Figure~\ref{ph-po2}(b) shows the behavior of phonon mode
at $\bf q \approx \frac{2}{3}$R as a function of the lattice constant.
It is seen that, with decreasing the lattice constant,
the phonon frequency rises very steeply in the vicinity of $a_{exp}$
(shaded area),
and so the imaginary phonon softening disappears for $a < a_{exp}$.
This is the reason why the phonon softening instability does not occur
at $a^0_{th}=3.335\AA$ ($V/V_{exp} = 0.98$) of the LDA+SOC,
as in the Verstraete's work \cite{Verst10},
but occurs at $a^0_{th}=3.411\AA$ ($V/V_{exp} = 1.05$)
of the GGA+SOC \cite{Min06}.
Moreover, since the $\alpha$ to $\beta$ transformation in Po occurs at 348 K,
one might have to take into account the thermal expansion ($\alpha_T$)
of the lattice constant\cite{Grabowski07,Hatt10}.
Considering $\alpha_T= 23.5$ $\mu m/m \cdot K$ of $\alpha$-Po
at 298 K~\cite{Lide}, the lattice elongation amounts to
$\sim 0.02$ $\AA$ at 300K, which is large enough to influence
the phonon softening instability.
In fact, the lattice constant of $\alpha$-Po at 298 K is reported
to be 3.366 $\AA$ ($V/V_{exp} = 1.01$) \cite{Lide}.
Then, without the SOC, $\alpha$-Po would be unstable to $\beta$-Po,
as indicated by the filled blue square in  Fig.~\ref{ph-po2}(b).

We now consider the superconducting properties
of the $\alpha$-Po~\cite{superconduc}.
In the Eliashberg theory,
the average EP coupling constant is given by
\begin{equation}
\lambda = 2\int_{0}^{\infty}d\omega\alpha^{2}F(\omega)/\omega
\end{equation}
with the Eliashberg function expressed as
\begin{eqnarray}
\alpha^{2}F(\omega)
& = & \frac{1}{N(E_F)}\sum_{\textbf{q}\nu}\sum_{\textbf{k},n,m}
|g_{\textbf{k}n,\textbf{k}+\textbf{q}m}^{\nu}|^{2} {}
\nonumber \\
& & \times~\delta(\epsilon_{\textbf{k}n})\delta(\epsilon_{\textbf{k}+\textbf{q}m})
\delta(\omega-\omega_{\textbf{q}\nu}).
\end{eqnarray}
Here $N(E_F)$ is the DOS at $E_F$, and $\omega_{\textbf{q}\nu}$ denotes
the frequency of the phonon mode $\nu$ with momentum $\textbf{q}$.
The EP matrix element
$g_{\textbf{k}n,\textbf{k}+\textbf{q}m}^{\nu}$
is obtained from $g_{\textbf{k}n,\textbf{k}+\textbf{q}m}^{\nu}
=\langle \textbf{k}n|\textbf{e}_{\textbf{q}\nu}
\cdot\nabla V_{KS}(\textbf{q})|
\textbf{k}+\textbf{q}m\rangle / \sqrt{2M\omega_{\textbf{q}\nu}}$,
where $\textbf{e}_{\textbf{q}\nu}$ is the polarization vector
and $\nabla V_{KS}(\textbf{q})$ is the gradient of the Kohn-Sham potential
with respect to the atomic displacements with the wave vector \textbf{q}.
Then the critical temperature $T_{c}$ is obtained
with the McMillan-Allen-Dynes formula \cite{Allen75} :
\begin{equation}
T_{c} = \frac{\omega_{\texttt{log}}}{1.20}
\texttt{exp}\bigg(-\frac{1.04(1+\lambda)}
{\lambda-\mu^{*}-0.62\lambda\mu^{*}}\bigg),
\end{equation}
where $\omega_{\texttt{log}}~\Big(\equiv \texttt{exp}\big(\frac{2}{\lambda}
\int_{0}^{\infty}\frac{d\omega}{\omega}\alpha^{2}
F(\omega)\texttt{ln}\omega\big)\Big)$ is the logarithmic average frequency
and $\mu^{*}$ is the effective Coulomb repulsion parameter.
The estimated superconducting parameters are listed in Table~\ref{Tc-table}.

\begin{table}[b]
\caption{DOS at $E_{F}$, $N(E_F)$, Debye temperature $\theta_{D}$,
and the superconducting parameters of $\alpha$-Po.
Two values of $T_{c}$ are for two different values
of $\mu^{*}$=0.10 and 0.13.
}
\begin{small}
\begin{tabular}{c|c c c p{0.4cm} p{0.4cm} c p{0.7cm} p{0.7cm}}
\hline \hline
& $V/V_{exp}$ & $N(E_F)$ & $\theta_{D}$ &
\multicolumn{2}{c}{$\lambda$} & $\omega_{\texttt{log}}$ &
\multicolumn{2}{c}{$T_{c}$} \\
& & (states/eV) & (K) & & & (meV) &
\multicolumn{2}{c}{(K)} \\
\hline
LDA+ & 1 & 0.63 & 103.7 &
\multicolumn{2}{c}{0.76} & 6.66 & 3.38, & 2.67\\
SOC & 0.98 & 0.60 & 101.5 &
\multicolumn{2}{c}{0.89} & 6.52 & 4.34, & 3.78\\
& 0.93 & 0.57 & 113.3 &
\multicolumn{2}{c}{1.04} & 6.52 & 5.57, & 5.10\\
\hline
LDA & 0.98 & 0.61 & 102.5 &
\multicolumn{2}{c}{0.95} & 6.43 & 5.04, & 4.25\\
& 0.93 & 0.57 & 110.6 &
\multicolumn{2}{c}{1.11} & 6.69 & 6.28, & 5.85\\
\hline \hline
\end{tabular}
\end{small}
\label{Tc-table}
\end{table}

Noteworthy in Table~\ref{Tc-table} is that $\alpha$-Po
at the ambient pressure ($V/V_{exp} = 0.98$)
would be a superconductor with $\lambda = 0.89$ and $T_c \sim 4 K$.
At the reduced volume ($V/V_{exp} = 0.93$),
both $\lambda$ and $T_{c}$ increase
further to 1.04 and $\sim$ 5 K, respectively.
With reducing the volume, $N(E_F)$ decreases,
while the Debye temperature $\theta_D$ increases.
The effect of the SOC on the superconductivity is found to be
not so large but detrimental.
For $V/V_{exp} = 0.93$, the SOC reduces $\lambda$ from 1.11 to 1.04,
and $T_{c}$ from $\sim$ 6 K to $\sim$ 5 K.
It is mainly because of the phonon hardening induced by the SOC,
as discussed in Figs.~\ref{ph-po1} and ~\ref{ph-po2}.
This behavior in Po is quite opposite to that in neighboring element Pb,
for which the SOC rather softens the phonons and increases
$\lambda$ by more than 40\% \cite{Verst08,Corso08,Heid10}.
This difference is thought to arise from the different Fermi surface
nature between two \cite{Verst08}.

\begin{figure}[t]
\includegraphics[width=8.5 cm]{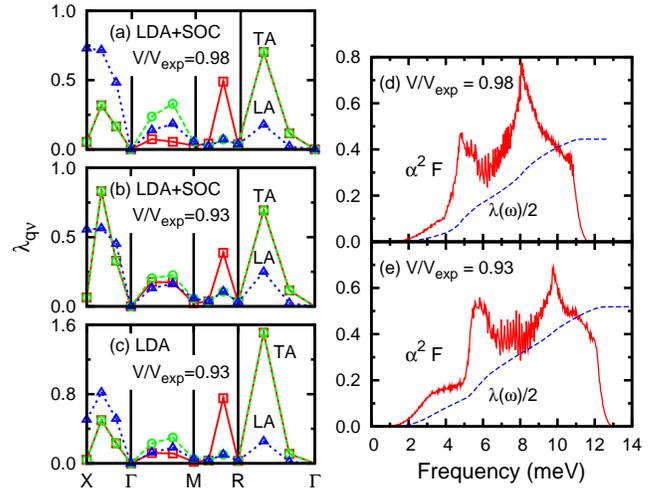}
\caption{(Color Online)
The \textbf{q}-dependent EP coupling constants of $\alpha$-Po:
(a) in the LDA+SOC for $V/V_{eq} = 0.98$,
(b) in the LDA+SOC for $V/V_{eq} = 0.93$,
(c) in the LDA for $V/V_{eq} = 0.93$.
The Eliashberg functions $\alpha^{2}F(\omega)$ with the SOC
at different volumes of $\alpha$-Po:
(d) in the LDA+SOC for $V/V_{eq} = 0.98$,
(e) in the LDA+SOC for $V/V_{eq} = 0.93$.
Integrated Eliashberg functions $\lambda(\omega)$ are shown.
}
\label{elph}
\end{figure}

To examine which phonon contributes largely to superconductivity,
we computed the \textbf{q}-dependent EP coupling constant
$\lambda_{\textbf{q}\nu}$.
As shown in Figs.~\ref{elph}(a)-(c),
$\lambda_{\textbf{q}\nu}$'s show peaks at \textbf{q}'s
having phonon softening anomalies
along $\Gamma$-X, M-R, and $\Gamma$-R directions.
Longitudinal X and transverse $\Gamma$-R phonon modes
have fairly large values of $\lambda_{\textbf{q}\nu}$.
We found above that $\lambda$ in $sc$-Po increases as the volume is reduced.
In view of the phonon hardening at reduced volume,
this behavior looks strange, at first glance.
The enhanced $\lambda$ can be explained by the
behavior of the Eliashberg function $\alpha^{2}F(\omega)$
in Figs.~\ref{elph}(d) and (e).
At the reduced volume, the maximum cut-off frequency is larger,
and $\alpha^{2}F(\omega)$ is enhanced at low frequency.
This leads to enhanced $\lambda$ and higher $T_c$.
As shown in Figs.~\ref{elph}(d) and (e), the integrated Eliashberg function
$\lambda(\omega)$ has a steep rise at about 4 and 5 meV
for $V/V_{exp} = 0.98$ and $V/V_{exp} = 0.93$, respectively.
This energy range corresponds to that of the transverse $\Gamma$-R
soft phonon mode (Fig.~\ref{ph-po2}),
suggesting that the superconductivity of $sc$-Po comes largely
from $\bf q \approx \frac{2}{3}$R phonon mode.
In the LDA scheme of Fig.~\ref{elph}(c),
the $\bf q \approx \frac{2}{3}$R soft phonon mode
produces the larger $\lambda_{\textbf{q}\nu}$,
while, in the LDA+SOC scheme of Fig.~\ref{elph}(b),
the $\bf q \approx \frac{2}{3}$R phonon mode hardened by the SOC reduces
$\lambda_{\textbf{q}\nu}$ by half.
This is the reason why the SOC is detrimental to the superconductivity in Po.

In conclusion, based on the phonon dispersion calculations,
we have explicitly demonstrated that the SOC is the origin of the stabilized
$sc$ structure of $\alpha$-Po.
We have also predicted that $\alpha$-Po would be a superconductor
with $T_c \sim 4$ K.
The soft transverse phonon modes at $\bf q \approx \frac{2}{3}$R
in $\alpha$-Po, which are greatly influenced by the SOC,
play an important role both
in the structural stability of the $sc$ phase and its superconductivity.
The experimental verification of the superconductivity
in $\alpha$-Po is demanded.

This work was supported by the NRF (No.2009-0079947, No.2011-0025237)
and the KISTI supercomputing center (No. KSC-2011-C2-36, KSC-2012-C3-09).



\begin{thebibliography}{99}
\bibitem{GGA}
    When the GGA (PBE) and PBEsol schemes are used,
	phonons are more softened to produce two imaginary phonon
	softenings along the M-R, and $\Gamma$-R directions,
	at both $a_{exp}$ and $a_{th}^{0}$ of the LDA+SOC.
    Note that $a_{th}^{0}$ of the PBEsol+SOC is 3.346 $\AA$.
\bibitem{superconduc}
    $\alpha$-Po has 33 known isotopes.
    Among those, the most abundant $^{210}$Po has the half-life of 138.3 days,
    and so the self-heating effect makes superconducting experiment difficult.
    Hence $^{209}$Po with half-life of 103 years, which is less vulnerable
    to the self-heating effect, would be more desirable for the
    superconducting experiment.
\end{thebibliography}
\end{document}